\documentclass[prd,
showpacs,showkeys,amsmath,amssymb,superscriptaddress]{revtex4}%

\usepackage{graphicx}
\usepackage{epstopdf}
\usepackage{array,amsmath,amssymb}

\begin{document}



\title{Infrared behavior of the Faddeev-Popov operator in Coulomb gauge QCD}

\author{Y.~Nakagawa} 
\affiliation{Research Center for Nuclear Physics, 
 Osaka University, Ibaraki, Osaka 567-0047, Japan}
\author{A.~Nakamura}
\affiliation{Research Institute for Information Science and Education, Hiroshima University, Higashi-Hiroshima 739-8521, Japan}
\author{T.~Saito}
\affiliation{Research Center for Nuclear Physics, 
 Osaka University, Ibaraki, Osaka 567-0047, Japan}
\author{H.~Toki}
\affiliation{Research Center for Nuclear Physics, 
 Osaka University, Ibaraki, Osaka 567-0047, Japan}



\begin{abstract}

We calculate the eigenvalue distribution of the Faddeev-Popov operator in Coulomb gauge QCD using quenched SU(3) lattice simulation.
In the confinement phase, the density of the low-lying eigenvalues increases with lattice volume, and the confinement criterion is satisfied.
Moreover, even in the deconfinement phase, the behavior of the FP eigenvalue density is qualitatively the same as in the confinement phase.
This is consistent with the fact that the color-Coulomb potential is not screened in the deconfined phase.

\end{abstract}




\maketitle


\section{Introduction}

One of the most challenging issues in elementary particle and nuclear physics is to understand the confinement of the quarks and gluons in quantum chromodynamics (QCD).
To understand the mechanism of confinement, there are several approaches in which topological objects are responsible for color confinement. Color monopoles in maximal Abelian gauge and center vortices in maximal center gauge are well-known examples \cite{GreensiteJ:PPNP51:2003,TokiH:PPNP45:2000}.
To clarify the mechanism of confinement, it is important to choose a proper gauge to extract the relevant degrees of freedom for color confinement.

The confinement mechanism in Coulomb gauge QCD has received a lot of attention recently.
Coulomb gauge is a physical gauge in the sense that there are no unphysical degrees of freedom.
Accordingly, Coulomb gauge is convenient for a variational approach and many attempts have been made to examine color confinement and hadron spectroscopy \cite{FeuchterC:PRD70:2004, ReinhardtH:PRD71:2005,SzczepaniakAP:PRD65:2001,BowmanPO:PRD70:2004,SzczepaniakAP:PRD69:2004,LigterinkN:PRC69:2004}.
It is firstly discussed by Gribov in '70s that the instantaneous interaction provides the long-range interaction \cite{GribovVN:NPB139:1978}, and this is further elaborated by Zwanziger recently \cite{ZwanzigerD:NPB518:1998}.
In Coulomb gauge, the time-time component of the gluon propagator can be decomposed into the instantaneous part and the noninstantaneous (polarization) part \cite{CucchieriA:PRD65:2001.1},
\begin{equation}
D_{00}(\vec{x},t)=I(\vec{x})\delta(t)+P(\vec{x},t).
\end{equation}
It has been found that $D_{00}$ is a renormalization-group invariant and, as a result, both $I$ and $P$ are separately renormalization-group invariants \cite{ZwanzigerD:NPB518:1998,BaulieuL:NPB548:1999,CucchieriA:PRD65:2001.1}.
Renormalizability of Coulomb gauge QCD was also studied within the Hamiltonian formalism \cite{ZwanzigerD:NPB518:1998,BaulieuL:NPB548:1999} and the Lagrangian formalism \cite{NiegawaA:PRD74:2006}.
Furthermore, Zwanziger showed that the color-Coulomb potential which is the instantaneous interaction energy between heavy quarks is stronger than a physical potential:
\begin{equation}\label{Zinequality}
V_{\textrm{Coul}}(R)\ge V(R).
\end{equation}
This inequality tells us that the necessary condition for the physical potential being a confining potential is that the color-Coulomb potential is also a confining potential, i.e., "no confinement without color-Coulomb confinement" \cite{ZwanzigerD:PRL90:2003}.

In SU(2) lattice calculations, it was reported that the instantaneous part of the gluon propagator, $I(\vec{k})$, is strongly enhanced at $\vec{k}=0$ \cite{CucchieriA:PRD65:2001,LangfeldK:PRD70:2004}.
Furthermore, the recent Monte Carlo simulations in the SU(2) and SU(3) lattice gauge theories showed that the color-Coulomb potential rises linearly with distance, and its string tension has $2\sim 3$ times larger value than that of the Wilson potential, which is an expected result from the Zwanziger's inequality \cite{GreensiteJ:PRD67:2003,NakamuraA:PTP115:2006}.
In addition, it was shown that an asymptotic scaling violation for the color-Coulomb string tension is weaker than that of the Wilson string tension \cite{GreensiteJ:PRD69:2004, NakagawaY:PRD73:2006}.

In the deconfinement phase, the static potential of a quark-antiquark pair is screened due to the screening effect \cite{KaczmarekO:PLB543:2002, KaczmarekO:PRD70:2004, NakamuraA:PTP111:2004, NakamuraA:PTP112:2004}.
In contrast, it has been shown by numerical simulations that the color-Coulomb potential is a confining potential even in the deconfinement phase, that is, the color-Coulomb potential is not screened above the critical temperature of the confinement/deconfinement phase transition \cite{GreensiteJ:PRD67:2003,NakamuraA:PTP115:2006}.
Thus the color-Coulomb string tension does not serve as an order parameter for the confinement/deconfinement phase transition.
This observation implies that the confinement is attributed to the instantaneous interaction in Coulomb gauge, whereas the confinement/deconfinement phase transition will be caused by the noninstantaneous interaction.

In Gribov-Zwanziger confinement scenario, the singular behavior of the color-Coulomb potential in the infrared region is governed by the near-zero modes of the Faddeev-Popov (FP) operator.
As Gribov pointed out, Coulomb gauge does not fix a gauge completely and the gauge configurations are restricted to the Gribov region where the FP operator is positive.
On the boundary of the Gribov region, so-called the Gribov horizon, the lowest eigenvalue of the FP operator vanishes.
Because of entropy considerations, a probability distribution gets concentrated near the Gribov horizon \cite{ZwanzigerD:NPB412:1994}.
The ghost propagator which is the expectation value of the inverse of the FP operator becomes singular in the infrared limit.
The color-Coulomb potential in the color-singlet channel is given by
\begin{equation}\label{CCP}
V_{\textrm{Coul}}(\vec{x}-\vec{y})\equiv g^2\textrm{Tr}[T^aT^b]\left\langle\int d^3z\mathcal{G}^{ac}(\vec{x}, \vec{z}; A^{\textrm{tr}})(-\nabla_{\vec{z}}^2)\mathcal{G}^{cb}(\vec{z}, \vec{y}; A^{\textrm{tr}})\right\rangle,
\end{equation}
where $\mathcal{G}$ is the Green's function of the FP operator and $\langle\cdot\rangle$ denotes an Euclidean expectation value.
$T^a$ ($a=1,..., 8$) are the generators of $\mathfrak{su}(3)$ Lie algebra.
The singular behavior of the ghost propagator in the infrared region leads to the long-range interaction of the color-Coulomb potential which is responsible for the color confinement.

Recently Greensite, Olejn\'{i}k and Zwanziger studied the spectrum of the FP operator in Coulomb gauge using SU(2) lattice gauge simulation.
The authors discussed the self-energy of an isolated quark and derived the necessary condition for the color confinement \cite{GreensiteJ:JHEP05:2005}.
It was shown that the FP eigenvalue density of the lowest modes becomes denser as the lattice volume increases and the necessary condition is satisfied in the infinite volume limit.

In this paper, we investigate the distribution of the FP eigenvalues in SU(3) lattice gauge simulations, and check whether the necessary condition for color confinement is satisfied or not.
In Sec. II we discuss the necessary condition for the confinement in Coulomb gauge and introduce the definition of the FP operator on a lattice.
Sec. III is devoted to show results of our numerical simulations. 
We discuss the confinement criterion also in the deconfinement phase.
In Sec. IV, we give conclusions.


\section{Color-Coulomb self-energy}

The Coulomb gauge Hamiltonian can be expressed as the sum of the gluonic part and the instantaneous part \cite{CucchieriA:PRD65:2001.1}:
\begin{eqnarray}\label{Hamiltonian}
H
&=&\frac{1}{2}\int d^3x\left((E_i^{a,\textrm{tr}}(\vec{x},t))^2+B_i^a(\vec{x},t)^2\right)+\frac{1}{2}\int d^3y\int d^3z\rho^a(\vec{y}, t)\mathcal{V}^{ab}(\vec{y},\vec{z}; A^{\textrm{tr}})\rho^b(\vec{z}, t).
\end{eqnarray}
Here $E_i^{a,\textrm{tr}}$ are the transverse components of the color electric field, $B_i^a$ the color magnetic field, $\rho^a(\vec{x},t)$ the color charge density.
The kernel of the instantaneous interaction is given by
\begin{equation}\label{kernel}
\mathcal{V}^{ab}(\vec{y},\vec{z}; A^{\textrm{tr}})\equiv\int d^3x\mathcal{G}^{ac}(\vec{y}, \vec{x}; A^{\textrm{tr}})(-\nabla_{\vec{x}}^2)\mathcal{G}^{cb}(\vec{x}, \vec{z}; A^{\textrm{tr}}),
\end{equation}
where $A_i^{a,\textrm{tr}}$ are the transverse components of the gluon field.
$\mathcal{G}$ is the Green's function of the FP operator $M^{ab}=-\partial_iD_i^{ab}=-\delta^{ab}\partial^2-gf^{abc}A_i^{c,\textrm{tr}}\partial_i$ whose expectation value $\langle\mathcal{G}^{ab}(\vec{x}, \vec{y}; A^{\textrm{tr}})\rangle=G(\vec{x}-\vec{y})\delta^{ab}$ is the ghost propagator.
The instantaneous interaction energy due to color charges originates from the longitudinal color electric field.
In this study, we focus on the relation between the instantaneous interaction and the spectrum of the FP operator, and we do not discuss the noninstantaneous interaction which may be relevant to the confinement/deconfinement phase transition.

The color-Coulomb self-energy for an isolated color charge, whose energy diverges in the infrared limit in a confining theory, is \cite{GreensiteJ:JHEP05:2005}
\begin{equation}\label{self-energy}
E_c=\textrm{Tr}[T^aT^b] g^2\langle \mathcal{V}^{ab}(\vec{x},\vec{x}; A^{\textrm{tr}})\rangle.
\end{equation}
The color-Coulomb self-energy is ultraviolet divergent in the continuum limit both in an Abelian and a nonAbelian gauge theories, and can be regularized by introducing the cutoff.
The interesting point is that the infrared divergence may exist in a confining theory in the infinite volume.

On a lattice, the FP operator is an $8V_3\times 8V_3$ sparse matrix ($V_3$ is the lattice 3-volume) and expressed in terms of SU(3) spatial link variables $U_i$ as 
\begin{eqnarray}\label{FPop}
M^{ab}_{xy}=&&\sum_{i}\mathfrak{Re}\textrm{Tr}\left[\{T^a, T^b\}\left(U_{i}(x)+U_{i}(x-\hat{i})\right)\delta_{x, y}\right.\nonumber\\
&&\qquad\qquad \left.\frac{}{}-2T^bT^aU_{i}(x)\delta_{y, x+\hat{i}}-2T^aT^bU_{i}(x-\hat{i})\delta_{y, x-\hat{i}}\right].
\end{eqnarray}
The Green's function of the FP operator can be expanded in terms of the eigenvectors $\phi^a_n(\vec{x})$ and the eigenvalues $\lambda_n$ of the FP operator;
\begin{equation}\label{greenfnc}
\mathcal{G}^{ab}(\vec{x}, \vec{y}; A^{\textrm{tr}})=\sum_n\frac{\phi^{\ast a}_n(\vec{x})\phi^b_n(\vec{y})}{\lambda_n}.
\end{equation}
From Eqs. (\ref{kernel}), (\ref{self-energy}) and (\ref{greenfnc}), we obtain
\begin{equation}
E_c=\frac{g^2C_D}{8V_3}\left\langle\sum_n\frac{F_n}{\lambda^2_n}\right\rangle.
\end{equation}
Here $C_D (>0)$ is the Casimir invariant for the representation $D$ and
$F_n$ the expectation values of the negative Laplacian in the FP eigenmodes,
\begin{equation}
F_n=\int d^3x\phi^{\ast a}_n(\vec{x})(-\nabla^2)\phi^a_n(\vec{x}).
\end{equation}
We define the normalized density of the FP eigenvalues
\begin{equation}
\rho(\lambda)\equiv\frac{N(\lambda, \lambda+\Delta\lambda)}{8V_3\Delta\lambda},
\end{equation}
where $N(\lambda, \lambda+\Delta\lambda)$ is the number of eigenvalues in the range $[\lambda, \lambda+\Delta\lambda]$.
The total number of the eigenvalues is $8V_3$ on a lattice and the FP eigenvalue density is normalized to 1.
Then we have
\begin{equation}
E_c=g^2C_D\int^{\lambda_{max}}_0 d\lambda\frac{\left\langle\rho(\lambda)F(\lambda)\right\rangle}{\lambda^2},
\end{equation}
where the upper limit of the integration $\lambda_{max}$ corresponds to the UV lattice cutoff.
In the Gribov-Zwanziger scenario the gauge configurations are restricted to the Gribov region, and therefore the lower limit of the integration is zero.
If the condition
\begin{equation}\label{condition}
\lim_{\lambda\to 0}\frac{\langle\rho(\lambda)F(\lambda)\rangle}{\lambda}>0
\end{equation}
is satisfied in the infinite volume limit, the color-Coulomb self-energy diverges in the infrared region.
This is the necessary condition for the color confinement \cite{GreensiteJ:JHEP05:2005}.
In this paper, we investigate whether the necessary condition for color confinement is satisfied in the quenched SU(3) lattice gauge simulation.

The FP eigenvalue density of the near-zero modes is closely related to the infrared behavior of the color-Coulomb potential.
From Eqs. (\ref{CCP}) and (\ref{self-energy}), the color-Coulomb self-energy can be expressed as
\begin{equation}\label{mspace}
E_c=V_{\textrm{Coul}}(\vec{x}-\vec{x})=\int\frac{d^3p}{(2\pi)^3}\tilde{V}_{\textrm{Coul}}(\vec{p})=\int_0^{\Lambda}\frac{d|\vec{p}|}{4\pi}|\vec{p}|^2\tilde{V}_{\textrm{Coul}}(|\vec{p}|).
\end{equation}
Here we introduce the ultraviolet cutoff $\Lambda$.
If the condition (\ref{condition}) is satisfied, the color-Coulomb self-energy diverges in the infrared region.
Accordingly, the right-hand side of Eq. (\ref{mspace}) diverges in the infrared limit.
It means that the color-Coulomb potential is more singular in the infrared region than the Coulomb potential $\tilde{V}(\vec{p})\sim 1/|\vec{p}|^2$.
Since the color-Coulomb potential provides an upper bound for the physical potential, the condition (\ref{condition}) is the necessary condition for the physical potential being a confining potential.

In the Abelian gauge theory (or at the zero-th order in the coupling), the FP operator is the negative Laplacian.
Thus the FP eigenfunctions are the plane waves and $\lambda=\vec{k}^2$.
By counting the number of states in momentum space, it is easy to show that
\begin{equation}\label{Abelian}
\rho(\lambda) = \frac{\sqrt{\lambda}}{4\pi^2},\qquad F(\lambda)=\lambda,
\end{equation}
in the infinite volume limit.
Obviously the necessary condition (\ref{condition}) is not satisfied in this case.

\section{Numerical simulations}

We calculate the FP eigenvalue density by the $SU(3)$ lattice gauge simulations in quenched approximation.
The lattice configurations are generated by the heat-bath Monte Carlo technique with the Wilson plaquette action (at $\beta=5.60 \sim 6.11$) and the Iwasaki improved action (at $\beta=2.605$), and we used 100 (60) thermalized configurations on $8^4 \sim 20^4$ $(24^4)$ at zero temperature
and $12^3 \times 6 \sim 24^3 \times 6$ at finite temperature.
In these simulations we adopt the iterative method to fix a gauge \cite{MandulaJE:PLB185:1987}.
In the iterative gauge fixing procedure, we minimize the functional
\begin{equation}
F_U[g]=\sum_{x,i}\textrm{Re}\textrm{Tr}\left(1-\frac{1}{3}g^{\dagger}(x)U_i(x)g(x+\hat{i})\right)
\end{equation}
with respect to the gauge transformation $g(x)$ and find the local minimum of $F_U[g]$.
The Hessian matrix associated with $F_U[g]$ is the lattice FP operator in Eq. (\ref{FPop}) and the local minima of the corresponding $F$'s define the Gribov region \cite{ZwanzigerD:NPB378:1992}.
Since the Hessian matrix is positive at a local minimum, the adopted method limits the lattice configurations to the Gribov region and the FP operator has no negative eigenvalues.
We used the LAPACK package to extract the whole eigenvalues of the FP operator on the $8^4$ lattice, while for larger lattice volumes, we used the ARPACK package to evaluate the lowest 1000 eigenvalues and corresponding eigenvectors of the FP operator because we are interested in the behavior of the low-lying FP eigenmodes. Since there are eight trivial zero modes corresponding to the spatially constant eigenvectors, we obtain the FP eigenvalue density from the remaining 992 eigenvalues.

\subsection{Full spectrum of the FP operator}

\begin{figure}[htbp]
\begin{center}
\resizebox{10cm}{!}{\includegraphics{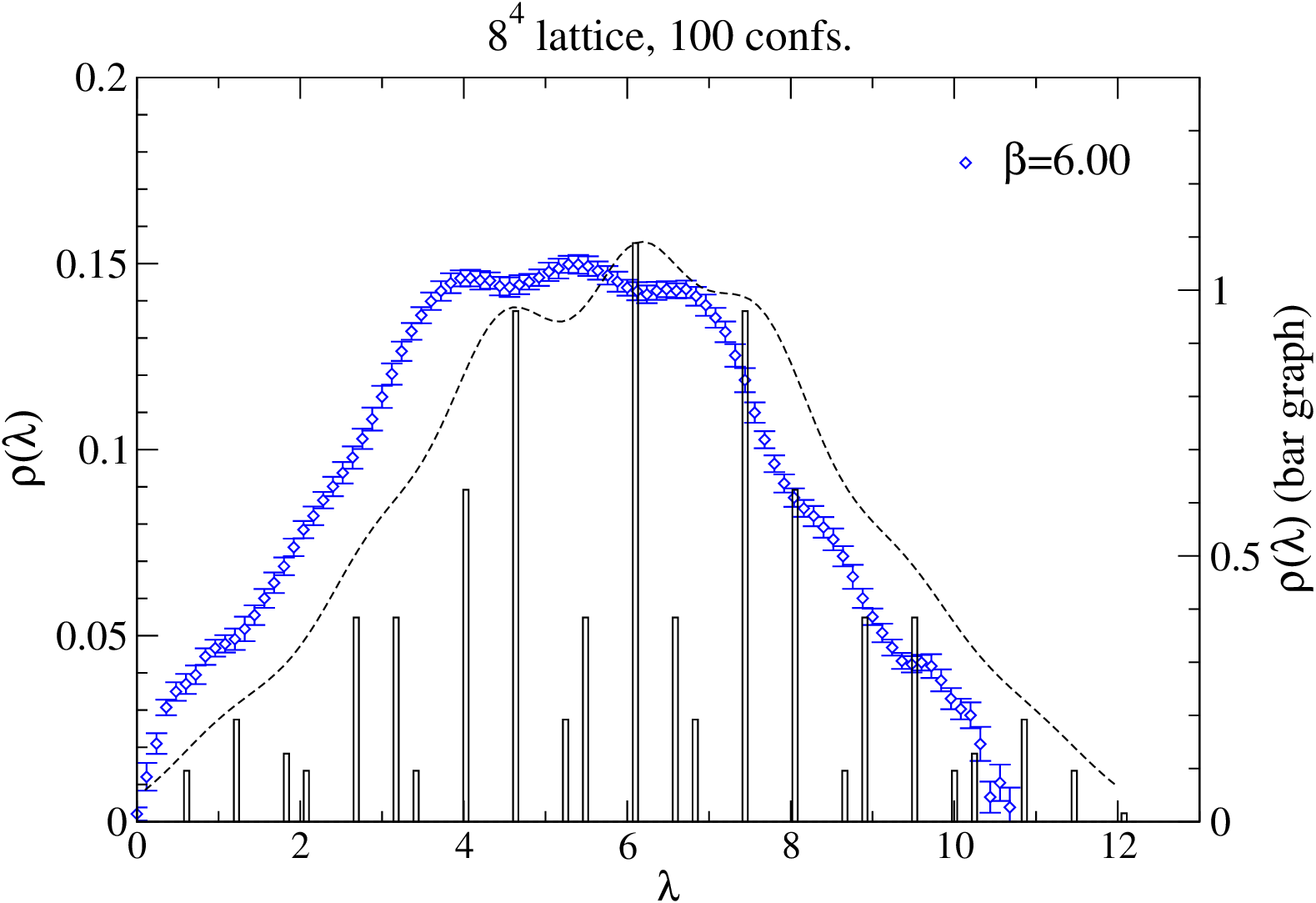}}
\caption{The FP eigenvalue density at $\beta=6.00$ on the $8^4$ lattice is shown by open diamond symbols with error bars as a function of $\lambda$. The vertical bars represent the FP eigenvalue density in the free field on the $8^4$ lattice and its scaling is displayed in the right-hand side of the figure. The dashed line is obtained by folding the vertical bars with normalized Gauss functions of a fixed width to smooth out the eigenvalue distribution for comparison with the interacting case.}
\label{EVBDEP-FULL}
\end{center}
\end{figure}

The full spectrum of the FP operator on the $8^4$ lattice at $\beta=6.00$ is displayed in Fig. \ref{EVBDEP-FULL} in lattice units, and the case of the free field, $U_i=I$, is also shown for comparison.
We show also a smooth eigenvalue distribution for the free field case by folding the vertical bars with normalized Gauss functions of a fixed width to smooth out the eigenvalue distribution for comparison with the interacting case (dashed curve).
We see that the whole eigenvalues of the FP operator shifts to lower values; namely, the number of the lowest eigenmodes is enhanced if the interaction turns on.

Furthermore, from this figure, we can see some bump structures, which correspond to the peaks in the case of the free field.
The FP operator for the free field is the negative Laplacian and the FP eigenvalues are, on a $L^4$ lattice,
\begin{equation}
\lambda=4\sum_{i=1}^3\sin^2\frac{n_i\pi}{L},\qquad n_i=0, 1,..., L-1.
\end{equation}
Thus, the FP eigenvalues degenerate and the eigenvalue density of the FP operator is the sum of the delta functions.
The degeneracy of the FP eigenvalues is lost and the peaks are broadened if the interaction turns on.
Eventually, the neighboring peaks overlap and the FP eigenvalues are distributed like those in the figure.
Therefore, the appearance of the bumps is due to the finite volume effect.

\begin{figure}[htbp]
\begin{center}
\resizebox{10cm}{!}{\includegraphics{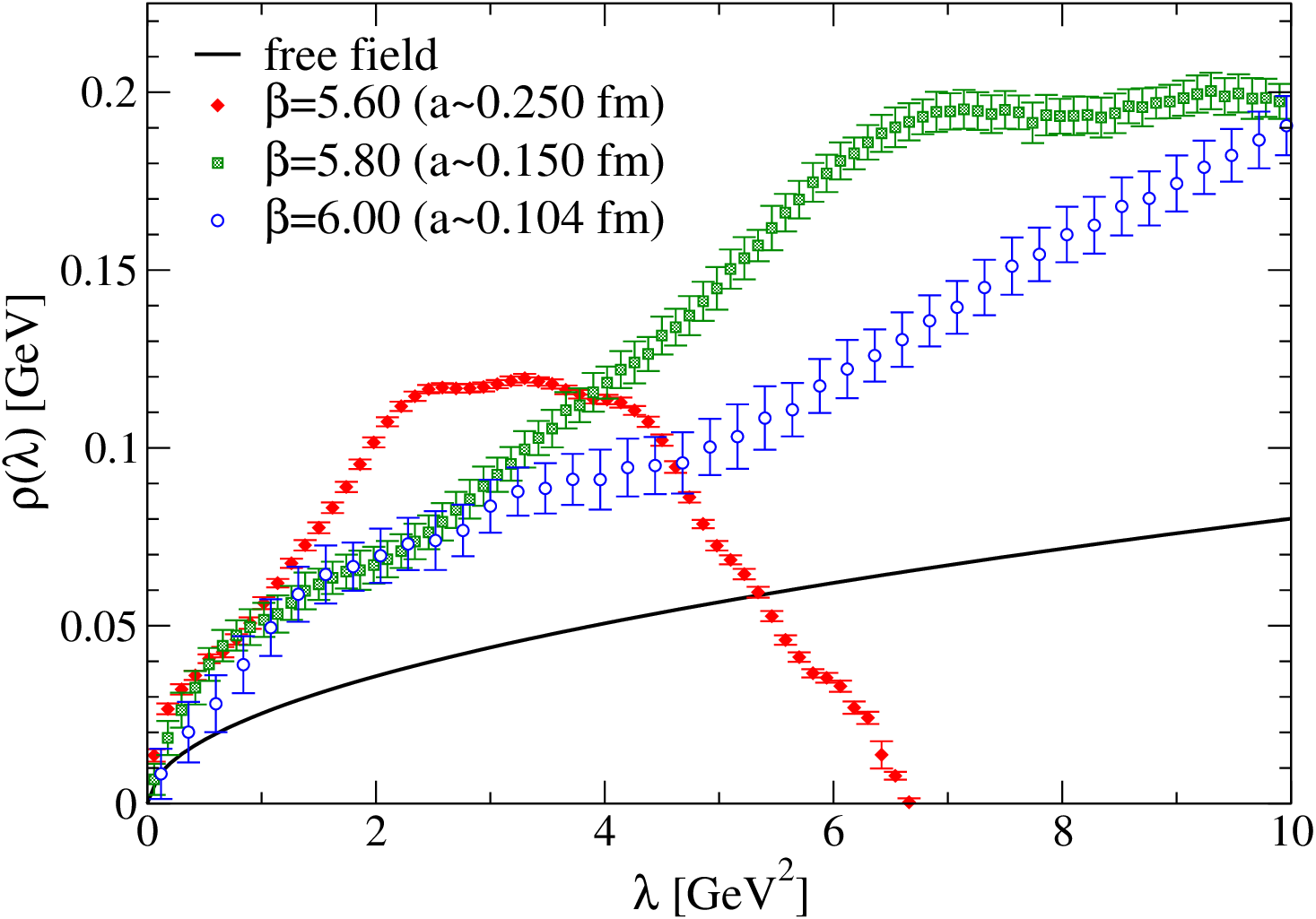}}
\caption{The FP eigenvalue density at $\beta=5.60 \sim 6.00$ on the $8^4$ lattice in physical units. The lattice couplings $\beta=5.60, 5.80, 6.00$ correspond to the lattice spacings $a\sim 0.250, 0.150, 0.104$ fm, respectively \cite{AkemiK:PRL71:1993}. The FP eigenvalues span a wider range as the lattice spacing decreases since $\lambda_{\textrm{max}}\sim 1/a^2$.}
\label{EVBDEP-PHYS}
\end{center}
\end{figure}

The FP eigenvalue density in physical units is shown in Fig. \ref{EVBDEP-PHYS}.
The three curves deviate from each other significantly above $\lambda\sim 1$ [GeV$^2$].
It means that the $\rho(\lambda)$ depends on the lattice cutoff seriously above $\lambda\sim 1$ [GeV$^2$].
However, we are interested in the behavior of the FP eigenvalue density near $\lambda=0$ and we will not discuss the cutoff dependence of the results anymore in this study.

\subsection{FP eigenvalue density with the improved action}

\begin{figure}[htbp]
\begin{center}
\resizebox{10cm}{!}{\includegraphics{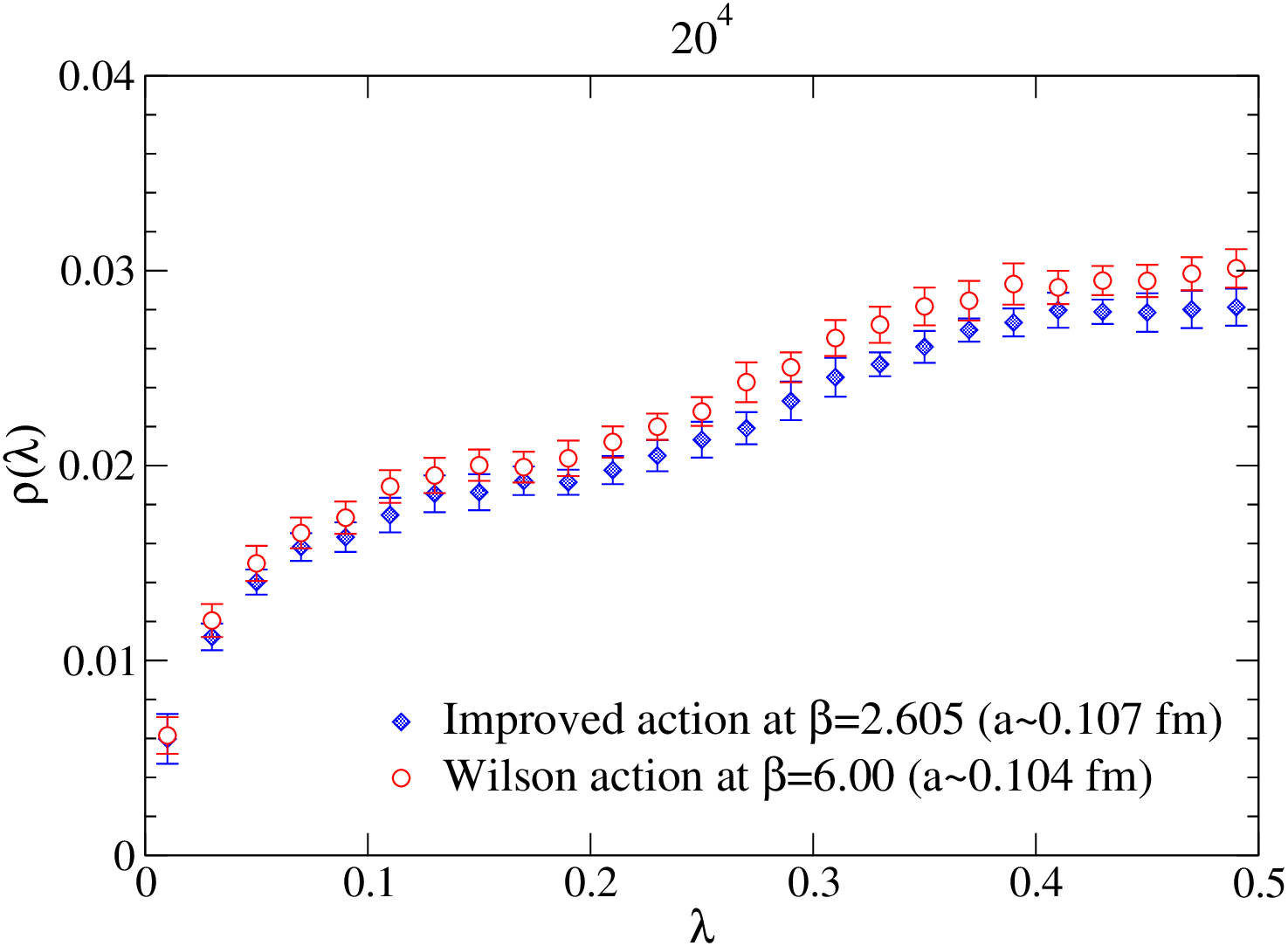}}
\caption{FP eigenvalue density on the $20^4$ lattice with the Wilson action and the Iwasaki action. The lattice coupling $\beta=2.605$ for Iwasaki action corresponds to the lattice spacing $a\sim 0.107$ fm, and $\beta=6.00$ for the Wilson action yields $a\sim 0.103$ fm.}
\label{Improved}
\end{center}
\end{figure}

In Fig. \ref{Improved} we plot the FP eigenvalue density on the $20^4$ lattice with the Iwasaki action \cite{IwasakiY:PRD56:1997}.
It is clear from this figure that the behavior of $\rho(\lambda)$ at small $\lambda$ with the Iwasaki action is similar to that of the Wilson action and there is no serious dependence of the FP eigenvalue density on the form of the lattice action.

\subsection{$\langle\rho(\lambda)F(\lambda)\rangle/\lambda$ in the confined phase}

\begin{figure}[htbp]
\begin{center}
\resizebox{10cm}{!}{\includegraphics{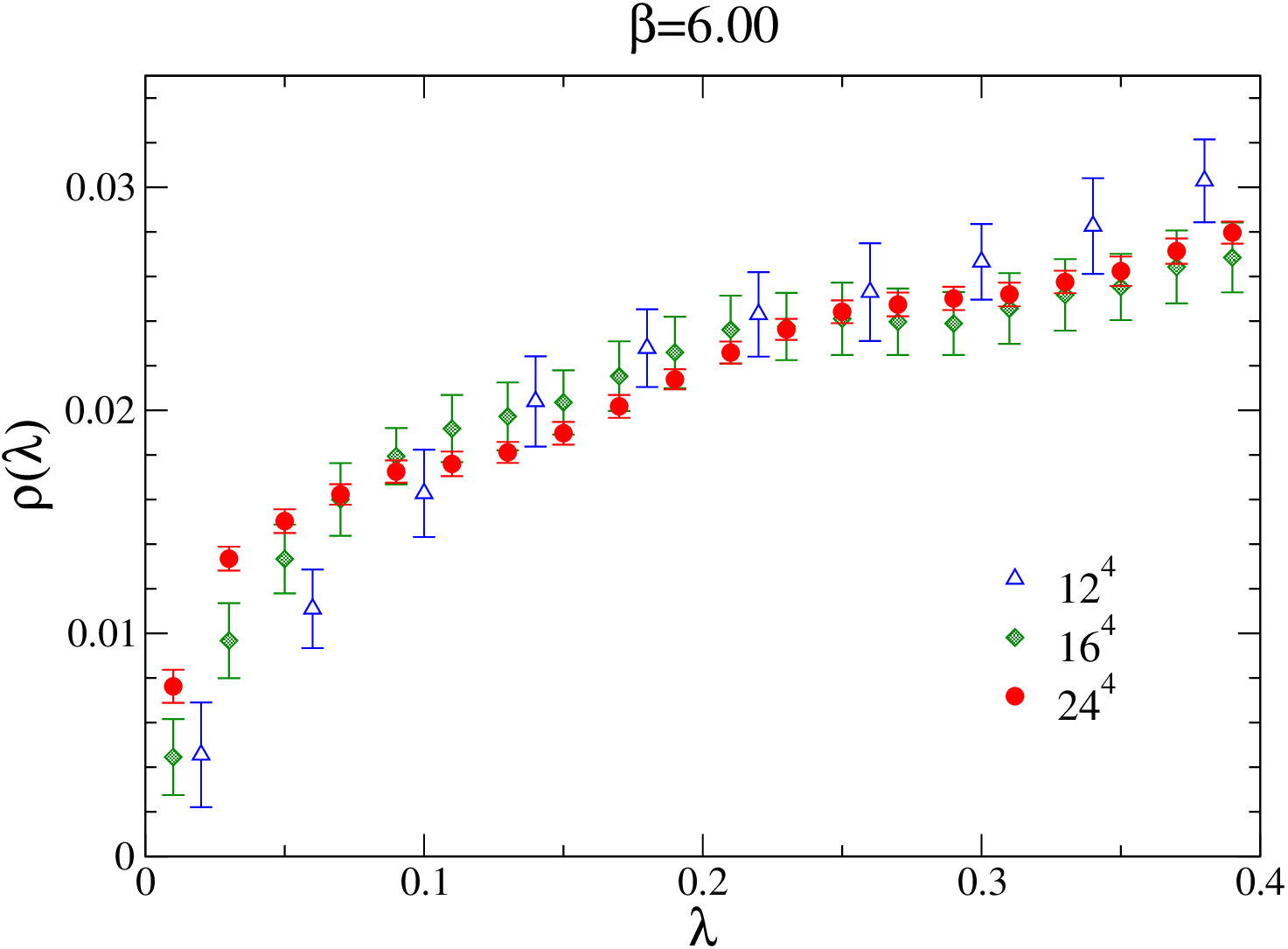}}
\caption{The FP eigenvalue density $\rho(\lambda)$ in the confinement phase on a variety of lattice sizes.}
\label{EVzero}
\end{center}
\end{figure}

\begin{figure}[htbp]
\begin{center}
\resizebox{10cm}{!}{\includegraphics{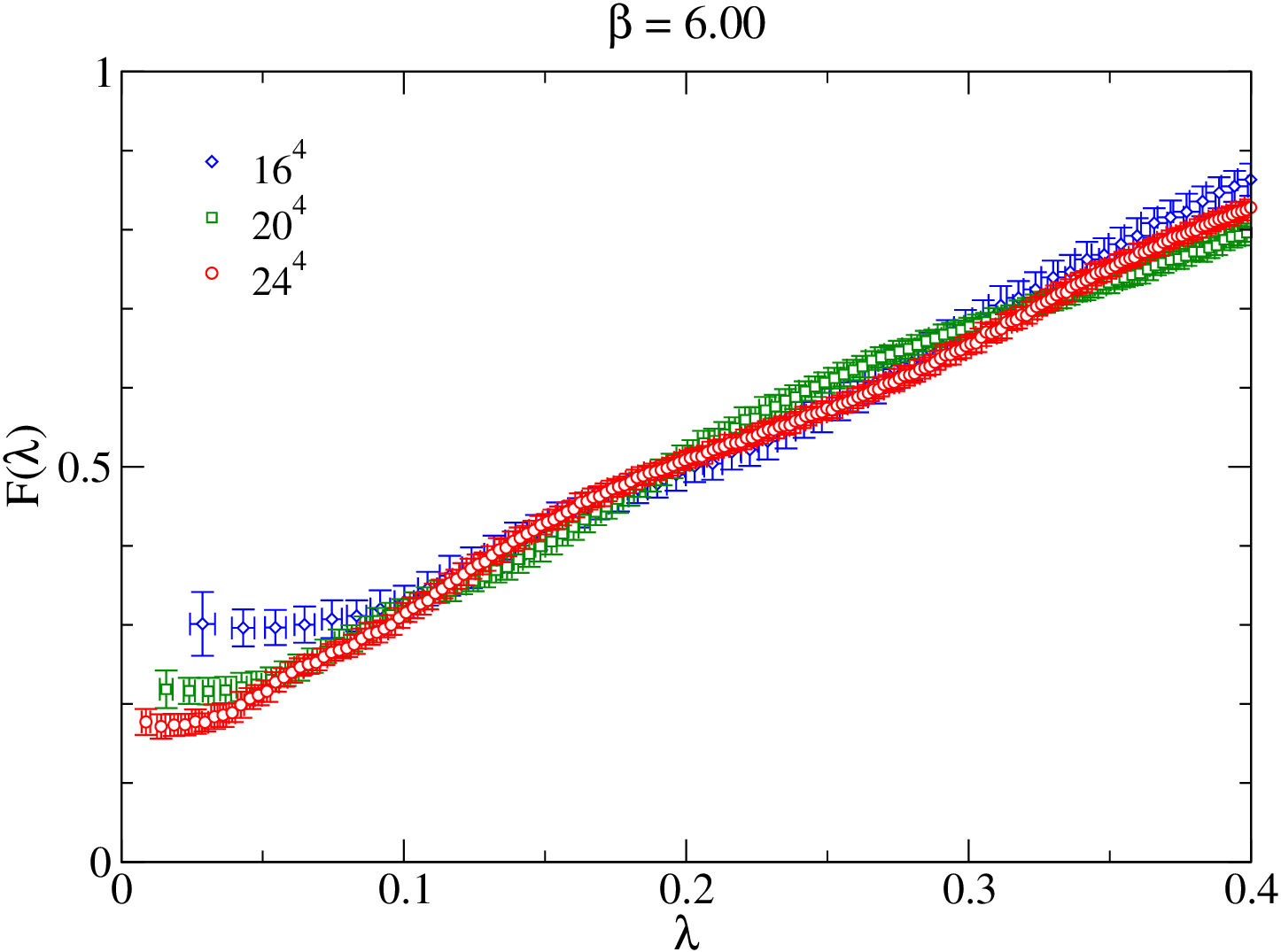}}
\caption{The average Laplacian $F(\lambda)$ in the confinement phase.}
\label{Fzero}
\end{center}
\end{figure}

Figures \ref{EVzero} and \ref{Fzero} show $\rho(\lambda)$ and $F(\lambda)$ at $\beta=6.0$ on $12^4 \sim 24^4$ lattice volumes.
We see that $\rho(\lambda)$ bends sharply as the lattice volume increases.
On the other hand, $\rho(\lambda)$ is almost saturated above $\lambda\sim 0.15$.
$F(\lambda)$ becomes flat at smaller value of $\lambda$ as the lattice volume increases.
From these figures, it seems that as $\lambda\to 0$ the FP eigenvalue density $\rho(\lambda)$ and the average Laplacian $F(\lambda)$ approach positive constants in the infinite volume limit; namely, the confinement criterion is satisfied in SU(3) Yang-Mills theory.

\begin{figure}[htbp]
\begin{center}
\resizebox{10cm}{!}{\includegraphics{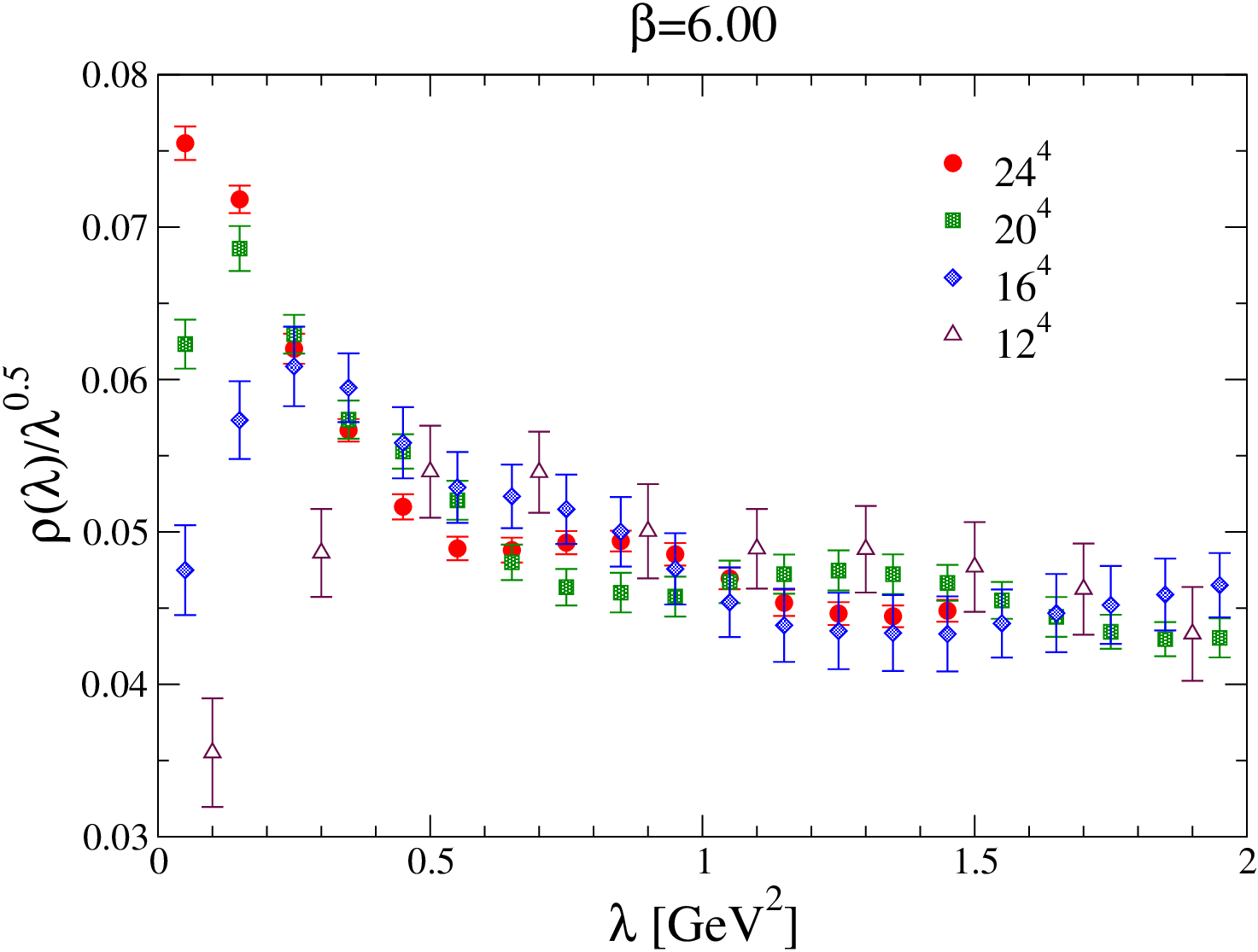}}
\caption{$\rho(\lambda)/\sqrt{\lambda}$ at $\beta=6.0$ on $12^4 \sim 24^4$ lattice volumes.}
\label{ENHANCE}
\end{center}
\end{figure}

To compare the behavior of $\rho(\lambda)$ in the nonAbelian theory with that in the Abelian theory, we plot $\rho(\lambda)/\sqrt{\lambda}$ in Fig. \ref{ENHANCE}.
In the Abelian theory it is constant because $\rho(\lambda)\sim\sqrt{\lambda}$ (see Eq. (\ref{Abelian})).
In the nonAbelian theory, we observe that $\rho(\lambda)/\sqrt{\lambda}$ is almost constant above $\lambda\sim 1$ [GeV$^2$].
It means that the qualitatative behaivor of $\rho(\lambda)$ in the nonAbelian theory is very similar to that of the Abelian theory at large $\lambda$.
In contrast, at small $\lambda$, $\rho(\lambda)/\sqrt{\lambda}$ is not constant and it seems to diverge as $\lambda\to 0$ in the infinite volume limit.
The FP eigenvalue density in the nonAbelian theory shows a completely different behavior at small $\lambda$ compared to that of the Abelian theory and we see the enhancement of the near-zero modes of the FP operator.
In the Gribov-Zwanziger scenario, these near-zero modes cause the color-Coulomb potential to be more singular in the infrared region than the simple pole. 

In the work given by Greensite et al., similar behaviors of $\rho(\lambda)$ and $F(\lambda)$ were observed in the SU(2) lattice gauge simulation.
However, they did not exclude the possibility that $\rho(\lambda)$ and $F(\lambda)$ vanish as $\lambda\to0$ and they gave the estimates
\begin{equation}
\rho(\lambda)\sim \lambda^{0.25}, \qquad F(\lambda)\sim \lambda^{0.38}
\end{equation}
by a scaling analysis.
Then they concluded that the necessary condition for the confinement is indeed satisfied in SU(2) Yang-Mills theory.
We have fitted the data on $24^4$ lattice with the functions,
\begin{equation}
\rho(\lambda) = c_1\lambda^p+c_2\sqrt{\lambda}, \qquad F(\lambda) = c_3\lambda^q + c_4\lambda,
\end{equation}
where we introduced the last terms $c_2\sqrt{\lambda}$ and $c_4\lambda$ which dominate the perturbative behavior at large $\lambda$.
The three-parameter fitting gives the exponents
\begin{equation}
p = 0.15(10), \qquad q = 0.29(4),
\end{equation}
with $\chi^2/ndf=1.60$ for $\rho(\lambda)$ and $\chi^2/ndf=0.763$ for $F(\lambda)$ respectively.

\begin{figure}[htbp]
\begin{center}
\resizebox{10cm}{!}{\includegraphics{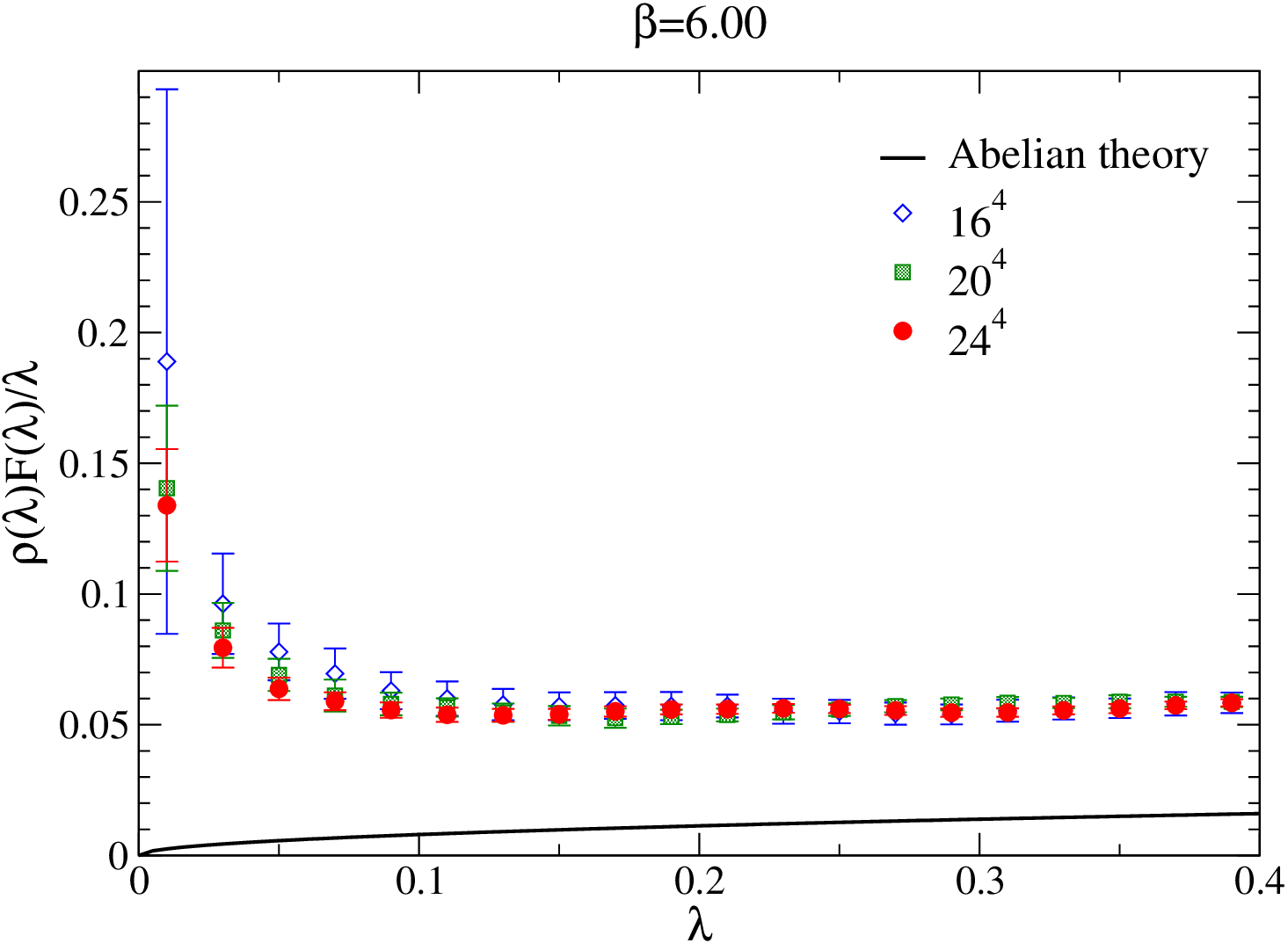}}
\caption{$\rho(\lambda)F(\lambda)/\lambda$ vs. $\lambda$ in the confinement phase.}
\label{FRHO}
\end{center}
\end{figure}

In Fig. \ref{FRHO} we plot $\rho(\lambda)F(\lambda)/\lambda$ as a function of $\lambda$.
As $\lambda$ approaches to 0, $\rho(\lambda)F(\lambda)/\lambda$ decreases for the free field (see Eq. (\ref{Abelian})) while increases for the interacting field.
In addition, the $24^4$ lattice simulation shows that $\rho(\lambda)F(\lambda)/\lambda$ at smaller $\lambda$ becomes flatter than the case of the $16^4$ lattice simulation.
From this figure, we expect that $\rho(\lambda)F(\lambda)/\lambda$ diverges or goes to positive constant, and it is unlikely that $\rho(\lambda)F(\lambda)/\lambda$ goes to zero as $\lambda\to 0$ in the infinite volume limit.
Therefore, we conclude that the color-Coulomb self-energy of an isolated color charge is infrared divergent in SU(3) gauge theory.

\subsection{$\langle\rho(\lambda)F(\lambda)\rangle/\lambda$ in the deconfined phase}

\begin{figure}[htbp]
\begin{center}
\resizebox{10cm}{!}{\includegraphics{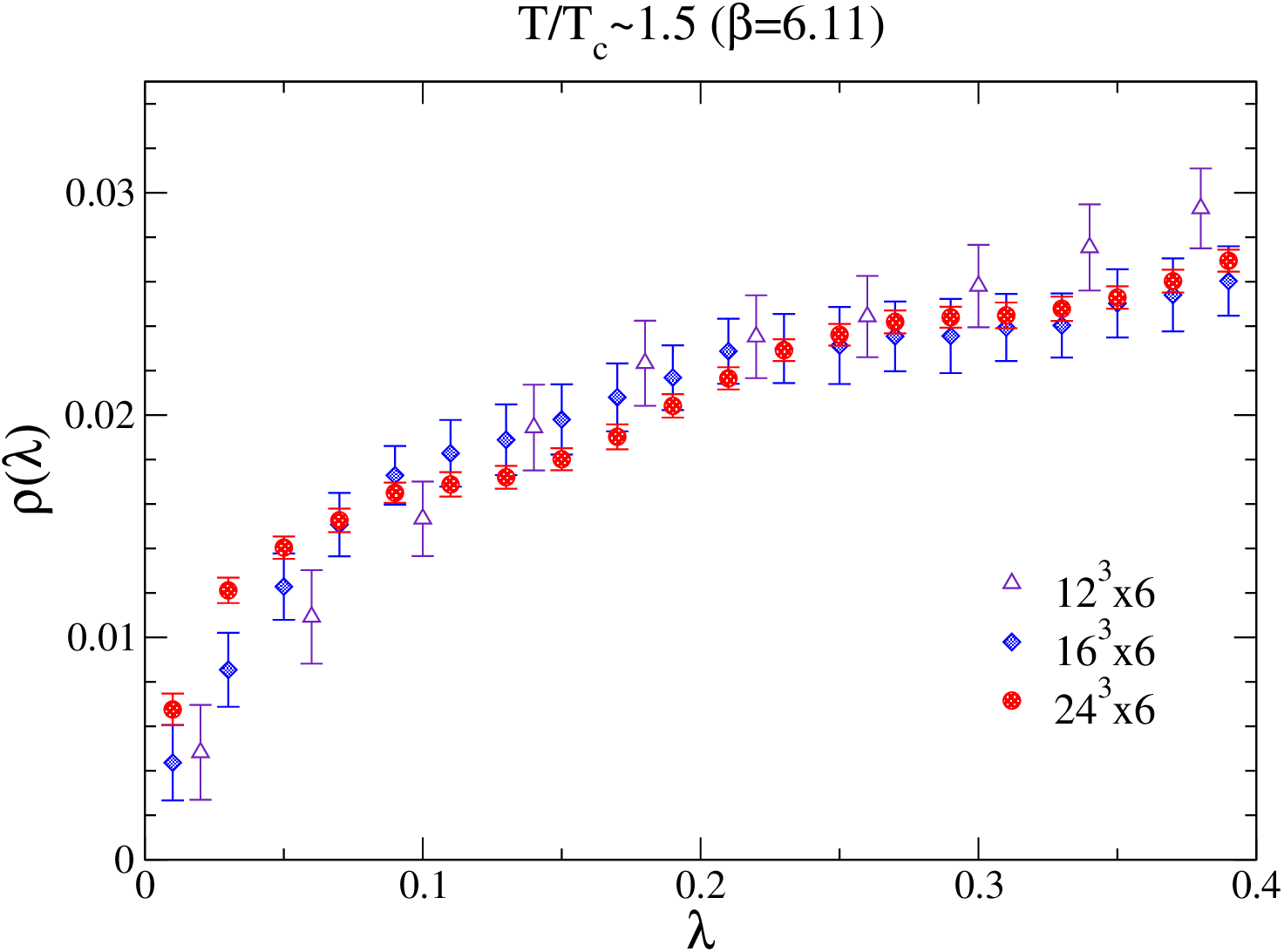}}
\caption{The FP eigenvalue density $\rho(\lambda)$ in the deconfinement phase.}
\label{EVfinite}
\end{center}
\end{figure}

\begin{figure}[htbp]
\begin{center}
\resizebox{10cm}{!}{\includegraphics{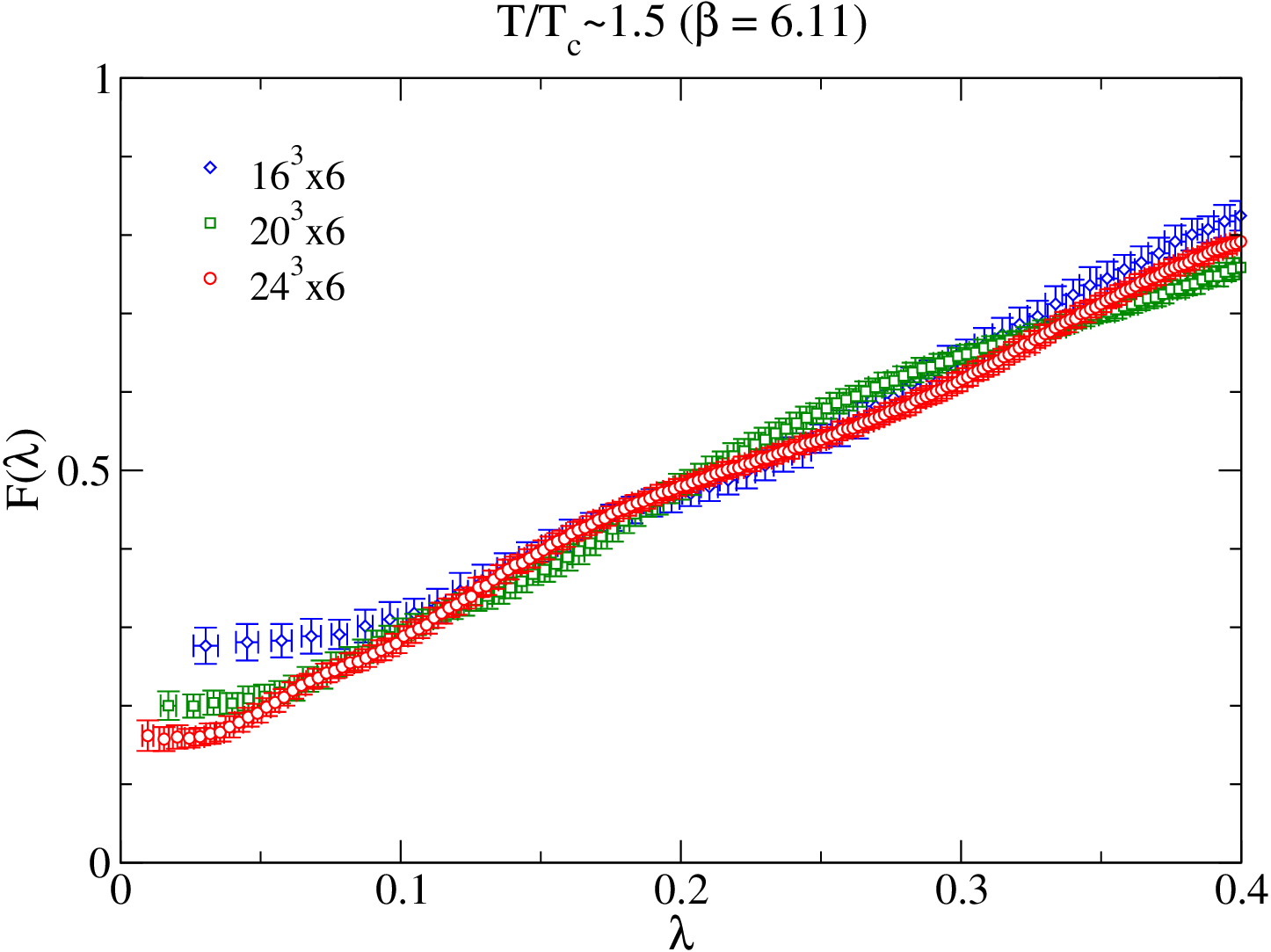}}
\caption{The average Laplacian $F(\lambda)$ in the decomfinement phase.}
\label{Ffinite}
\end{center}
\end{figure}

\begin{figure}[htbp]
\begin{center}
\resizebox{10cm}{!}{\includegraphics{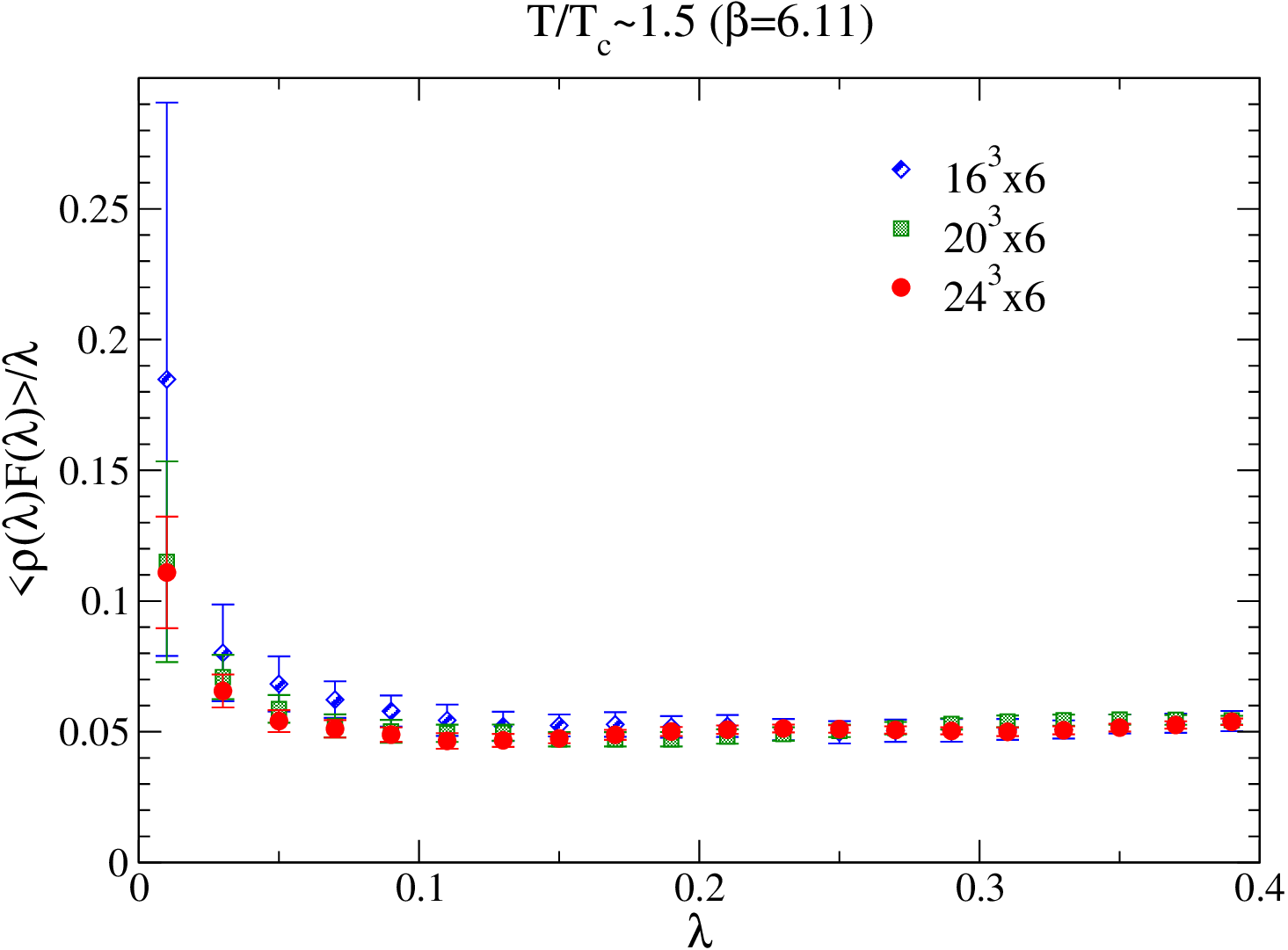}}
\caption{$\rho(\lambda)F(\lambda)/\lambda$ vs. $\lambda$ in the deconfinement phase.}
\label{FRHOLAMfinite}
\end{center}
\end{figure}

$\rho(\lambda)$, $F(\lambda)$ and $\rho(\lambda)F(\lambda)/\lambda$ in the deconfinement phase on $12^3\times 6 \sim 24^3\times 6$ lattice volumes are displayed in Figs. \ref{EVfinite}, \ref{Ffinite} and \ref{FRHOLAMfinite}.
The simulations are carried out at $\beta=6.11$ which corresponds to $T/T_c\sim 1.5$.
Here $T_c$ is the critical temperature of the confinement/deconfinement phase transition.
These figures show that there are no drastic changes of these behaviors in the deconfinement phase.
This is consistent with the facts that the color-Coulomb potential rises linearly with distance even in the deconfinement phase and the color-Coulomb string tension does not serve as an order parameter for the confinement/deconfinement phase transition.
We conclude that the necessary condition for color confinement is satisfied even in the deconfinement phase.

This result may not be a surprising result, since in Coulomb gauge the FP operator is a purely spatial quantity and does not depend on time explicitly.
On the other hand, the time extent of a lattice determines temperature.
Therefore, the FP operator is insensitive to temperature and it is natural that the spectrum of the FP operator does not show a critical behavior.

\section{Conclusions}

We have calculated the eigenvalue distribution of the FP operator in Coulomb gauge using quenched SU(3) lattice gauge simulations.
In the confinement phase, we observe the accumulation of the near-zero eigenvalues of the FP operator at large lattice volumes.
Moreover, the lattice simulations reveal that the average Laplacian approaches positive constant as $\lambda\to0$.
We conclude that the confinement criterion is satisfied in the SU(3) gauge theory.
This supports the Gribov-Zwanziger confinement scenario.
The results we obtained are qualitatively consistent with those of the SU(2) lattice simulation carried out by Greensite et al.

The near-zero modes of the FP operator survive above the critical temperature, and the behaviors of the FP eigenvalue density and the average Laplacian in the deconfinement phase are qualitatively the same as in the confinement phase.
Accordingly, the confinement criterion is satisfied even in the deconfinement phase in SU(3) gauge theory.
It is not surprising that the spectrum of the FP operator is insensitive to temperature, because the FP operator is a spatial quantity.
We note that the criterion is not a sufficient condition but a necessary condition for the confinement; namely, the color-Coulomb energy is not the ground state energy but the excited state energy of color charges \cite{ZwanzigerD:PRL90:2003}.
If we take the noninstantaneous interaction into account when discussing in the deconfinement phase, the energy of an isolated color charge will be finite in the infrared limit and the isolated color charge can exist.

As we have shown in this paper, the spectrum of the FP operator does not change drastically above the critical temperature.
This would indicate that confining features survive even in the deconfinement phase.
Actually, it is known that the spatial Wilson loop which is a gauge invariant quantity shows the area law behavior above the critical temperature.
Therefore, we expect that further studies in Coulomb gauge provide insight into the understanding of the strongly correlated quark-gluon plasma.

The color-Coulomb potential can be obtained by calculating the FP eigenvalues and eigenvectors.
It is valuable to see whether the lowest eigenmodes of the FP operator produce the linearly rising behavior of the color-Coulomb potential for large quark separations.
We address this issue in our future investigation.

\section{Acknowledgments} 

The simulation was performed on an SX-5(NEC) vector-parallel computer 
at the RCNP of Osaka University. 
We appreciate the warm hospitality and support of
 the RCNP administrators.
This work is supported by Grants-in-Aid for Scientific Research from
Monbu-Kagaku-sho (Nos. 13135216 and 17340080).

\bibliographystyle{h-physrev4}
\bibliography{../forQCD}

\end{document}